\documentclass[12pt]{article}		%% Font: Times New Roman, 12 pt. %%
\usepackage{fancyhdr}
\usepackage{multicol}
\usepackage{latexsym}
\usepackage{graphicx}
\usepackage[T1]{fontenc}			%% Allows for additional special characters
\usepackage{setspace}				%% To set line spacing manually %% 
\usepackage{endnotes}				%% For if journal uses endnotes %%
\usepackage{marvosym}				%% Necessary for using the @ symbol %%
\usepackage{seqsplit}				%% Makes it so that long words will be automatically parsed. Useful for web addresses %%
\usepackage[hyphens]{url}

\usepackage[round,semicolon,authoryear]{natbib}	%% Find-Replace "\citealt" for \cite if journal wishes parentheses after AuthorName to be eliminated %%
\usepackage{authblk} % For author affiliations

\begin{document}
	
%\title{Defining Autonomous Weapon Systems: A definitional schema for a classificatory framework}
\title{\vspace{-4.0cm}Defining Autonomous Weapon Systems: A schematic framework for refined purpose-driven definitions}
%\title{Defining Autonomous Weapon Systems}
\author[1,2]{\textbf{Bj\"{o}rn Lundgren},\thanks{Initial ideas for this work were presented by BL in ``A Definitional Framework of Autonomous Weapon Systems'' at IACAP 2024, July 8-10, Eugene, Oregon, USA, as well as in NW's articles ``Autonomous Weapon Systems: A clarification'', \textit{Journal of Military Ethics}, 22.1 (2023) and ``Stop Saying `AI'\thinspace'', \textit{Philosophy \& Technology}, 39.2 (2026). Both authors contributed to the drafting and revision of the entire work, and are jointly responsible for all arguments and conclusions. Authors listed alphabetically.} bjorn.lundgren@fau.de} % arXiv username: bl-arxiv :: email: bjorn.lundgren@iffs.se
\author[3,4,5]{\textbf{Nathan G. Wood}, nathan.wood@tuhh.de}
\affil[1]{Centre for Philosophy and AI Research, Friedrich-Alexander University Erlangen-Nuremberg, Germany}
\affil[2]{Institute for Futures Studies, Sweden}
\affil[3]{Institute of Air Transportation Systems, Hamburg University of Technology, Germany}
\affil[4]{Ethics + Emerging Sciences Group, California Polytechnic State University San Luis Obispo, United States of America}
\affil[5]{Center for Environmental and Technology Ethics -- Prague, Czechia}
\date{(Initial Draft)\\ \today}

\maketitle	

%\pagebreak

\renewcommand*\abstractname{Abstract\hfill}
\begin{abstract}

The literature on autonomous weapon systems (AWS) includes many proposed definitions for these (types of) systems, with core disagreements between scholars, governments, and international regulatory bodies. In this article, we present and critique prevailing (classes of) definitions, arguing that what is needed is not a singular definition, but rather a definitional framework which allows one to classify different sorts of AWS within the same conceptual family. Recognizing we cannot foresee all future AWS, we show how variations on a broadly accepted definition can be used to produce a definitional schema which can be easily modified in order to classify various concepts belonging to the class of AWS.

\bigskip
	
\noindent\textbf{Keywords:} \emph{Autonomous Weapon Systems; Autonomy; AI; Military Ethics; Arms Regulation; AI Governance; Responsible Innovation}		%% Usually between 3 and 5 keywords are expected. %%

\end{abstract}

\pagebreak

\section{Introduction}\label{sec_intro}

% Intro
As of writing this, it has been over a decade since regulatory discussions around autonomous weapon systems (AWS) began in Geneva under the auspices of the United Nations Convention on Conventional Weapons (CCW),\footnote{Fully, the ``Convention on Prohibitions or Restrictions on the Use of Certain Conventional Weapons Which May be Deemed to be Excessively Injurious or to Have Indiscriminate Effects''. For a brief historical overview of the work at the CCW, see \url{https://dig.watch/processes/gge-laws}, accessed July 31, 2026. For the rolling text from those efforts, see \cite{gge2025rolling12052025}.} and ethical discussions have been ongoing far longer.\footnote{Early discussion of ethical and doctrinal aspects of autonomous weapons can be found in, e.g., \cite{usnavy2005laws,sharkey2007automated,sharkey2008grounds,sparrow2007killer,lin2008autonomous}.} In that time, many definitions of AWS have been put forward,\footnote{E.g., \cite{roff2013killing}, pp. 353--354; \cite{icrc2014autonomous}, p. 5; \cite{williams2015defining}; \cite{boothby2016weapons}, pp. 247--257; \cite{caron2020defining}; \cite{icrc2021positiona}, p. 1; \cite{taddeo2022comparative}; \cite{usdod2023directive3000.09}, p. 21. See \cite{wood2023awsclarification,pacholska2024autonomous} for broader discussion of definitions and definitional worries.} and while some have gained wider acceptance, there is still no broadly agreed view. Moreover, the range of existing definitions all admit weaknesses in need of addressing. This article seeks to remedy this by developing not a \textit{singular definition} of AWS, but rather a \textit{conceptual definitional framework} for defining specific (types of) autonomous systems used in various military roles. In doing so, the article pays heed to a core reality of autonomy in systems, namely that it can be leveraged for a variety of purposes, with a variety of technical, operational, legal, and ethical challenges, and with various attendant technical, operational, legal, and ethical approaches for mitigating those challenges.\footnote{\cite{scharre2023four}, p. 3; \cite{wood2026stop}.} Thus, given the need for specificity, the article builds toward a modular definitional schema where scholars, policy-makers, and defense professionals can ``plug in'' definitional components as needed, befitting both the system they are considering and their definitional needs. This latter concern is of relevance given that not all definitions serve the same purpose, and in a domain as politically charged and with as significant of consequences as this, being able to craft purpose-built definitions for specific applications holds immense value. 

% Sectioning
To develop these arguments, the article begins (Section \ref{sec_definitions}) by canvassing general points on definitions, laying out the methodological assumptions guiding this work and tying these to the underlying purpose of the article. With these in place, Section \ref{sec_existing-views} then presents and critiques two strands of definition one can find in the literature from scholars, states, and international bodies, what we call the ``standard view'' and the ``demanding view''. Building from these critiques, Section \ref{sec_def.-aws} lays out base definitions of both ``autonomous systems'' and ``autonomous weapon systems'', stressing that these are \textit{base definitions}. That is, they are not to be treated as complete, as demarcating every relevant feature of some system, or as not needing further refinement. Indeed, Section \ref{sec_conceptual-framework}, which forms the central result of this work, serves as an extended presentation of various core concepts which should be further indicated in the definitions of particular autonomous systems in the military (and in other domains as well). This section concludes with general recommendations for constructing purpose-built definitions which include all relevant terms for a given system and use-case. %The article concludes (Section \ref{sec_conclusion}) with a broad defense of the need for specificity in technological, operational, legal, regulatory, and ethical discussions around autonomous weapons, and reiterates that the definitional schema developed here can, with suitable alterations, be used in domains other than the military where high-level technologies and capabilities (such as autonomous systems, artificial intelligence, agentic architectures) allow for numerous sub-systems to be built which raise distinct concerns. 
The article concludes (Section \ref{sec_conclusion}) with a broad recapitulation of the need for a pragmatic and purpose-oriented definitional schema, rather than an intellectualized pursuit of ``being right'' at the cost of progress.

% Two preliminary points: 
% 1. The purpose of this work is to add clarity but not argue for any further conclusions (e.g., concerning the permissibility of some (types of) systems)
Before moving onto the arguments themselves though, there are two preliminary points to be addressed. First, it is worth clarifying and stressing at the outset that the purpose of this work is to add clarity to the ongoing debates surrounding autonomous (weapon) systems and to reduce the risks of scholars and practitioners ``talking past one another''. It is not our aim to weigh in on any particular debates themselves though, nor to advocate for the benefits or risks of certain systems, their permissibility, or any related concern. That work is highly important, and it requires foundational definitions to be well-articulated, and it is this latter concern only which is addressed by the work here. Our aim is thus to develop a non-prejudicial, generally acceptable, and practically useful definitional schema, but not to advocate for anything beyond that.

% 2. The discussion here is specific to autonomous \textit{weapon} systems, but the definitional schema can be applied (with appropriate modification of pluggable modular terms) for technologies in other domains as well; it is thus compementary to efforts such as the recent work by \cite{wood2026stop}, especially the ``template for policy writing'' presented therein (p. 23)
Second, though we are discussing autonomous weapon systems in particular, the general points surrounding definitions and our schematic for constructing modular definitions which can be tailored to specific systems and use-cases holds general import. Thus, provided appropriate modifications are made to the list of pluggable modular terms, the overall schema and its motivating points can be ported to other technologies and domains of application. This is complementary to recent work such as that from \cite{wood2026stop} which argues for the need for specificity, and Wood et al.'s ``template for policy writing'' on AI\footnote{\cite{wood2026stop}, p. 23.} fits neatly with the overall points developed here. 

\section{What's In a Definition?}\label{sec_definitions}

% Definitions can be ``scientific'' or ``engineering'' in nature, aiming at splitting nature at the joints or at solving some human issue; we take a pragmatic approach that tries to balance both these concerns, but which errs on the side of engineering when conflicts arise between these methodological approaches; briefly touch on the point about ``completeness'' from first draft: ``Preferably, such a definitional framework should be complete. However, since completeness is epistemically impossible (e.g., because we foresee all future technological development, nor can we determine what classifications that are currently needed in all possible context). Hence, instead of a definitional framework, I will provide a definitional schema, which can be easily modified to adapt to any interesting class of AWSs.''
Not all definitions are created equal. If asked for a definition of a ``yellow lab'', one could answer that this is an organism of the animal kingdom, of the chordata phylum, mammal class, carnivora order, canidae family, from the genus canis, species canis lupus, subspecies canis lupus familiaris, and breed Labrador Retriever. But also, since it is yellow, one must add that on top of these various taxonomic details, it is a specific set of characteristics regarding the MC1R, Agouti, and CBD103 genes that make it a ``yellow'' lab. Alternatively, one could define this as a dog that is traditionally used for retrieving hunted fowl. Both are fine definitions... in the right setting. 

At a basic level, definitions can be built in what we might think of as being molded to science or engineering. A ``scientific'' definition may be seen as aiming to carve nature at the joints, capture the \textit{true} state of things in some domain, or otherwise direct individuals toward knowledge or some ontological facts.\footnote{\cite{scriven1958definitions,caws1959functions,michie2019creating}.} An ``engineering'' definition, on the other hand, is one which is aimed at \textit{being useful}, at solving some set of problems humans have and providing a ``good enough'' approximation for some given aims.\footnote{\cite{dym1994engineering,hossain2018synthesis,sillitto2019systems}. See also \cite{hibberd2019scientific} for useful comparative discussion of definitional types.} There will be crossover between these two sorts of definitions as well, determined by use cases and the purpose of asking for a definition in the first place. For example, in most instances, the first definition of ``yellow lab'' above could be seen as almost purely scientific in nature, admitting little ``usefulness''. However, if a geneticist is asking for the definition, then the taxonomy and particular genes involved in a yellow lab's being yellow may be highly relevant for a particular aim or inquiry the geneticist has in mind. The reason for asking can thus determine the quality of a definition. 

In this work, we take a broadly pragmatic approach that tries to balance these concerns, but errs on the side of viewing definitions through an engineering lens when conflicts might arise. This is partly because the definitions of man-made artifacts are bound to be at some point arbitrary anyway, and because the domain we have in mind, emerging technologies in the military domain, is one where definitions have extreme functional import for debates. There is thus value in looking at a definition not primarily as a marker of ``what is'', but rather as a tool for facilitating discussion and debate, effective regulation, common understanding, etc. Moreover, given that technologies continuously develop, no definition of a technical artifact is apt to ever be properly complete, undermining the possibility for developing fully satisfactory scientific definitions. 

% Definitions can serve a variety of purposes (driving scientific/philosophical debate, grounding technical documentation and training, setting agreed terms for regulation/norms); these different purposes will call for slightly different expressions of definitions (hearken to literature on explanations here); being bullish on any particular definition makes one apt to overlook critical shortcomings one's preferred definition may have in related but separate domains
As noted above, definitions can also serve a variety of purposes, grounding scientific or philosophical discussions, setting agreed terms for regulation/norms, or establishing enough understanding for technical documentation and training processes to be established. These different purposes will call for slightly different expressions of definitions, or for definitions which are more or less closely tied to what we take to be the ground truth. For example, Isaac Newton defined numerous terms which, in classical mechanics and everyday life, are very useful. However, those same definitions do not account for quantum or relativistic effects, and thus are technically wrong, scientifically speaking. On the other hand, quantum physics is not exactly the most practical discipline, and so it would be pragmatically wrong, from an engineering perspective, to rely on it for everyday engineering or design purposes. The simple fact is that a particular definition may be ``good'' or ``bad'' for a variety of reasons, and a central determinant of a definition's ``goodness'' relates to what the definition \textit{is for} in some instance. 

Being overly bullish on the priority or value of any one particular definition makes one apt to overlook its shortcomings and the value of potential competitors for given tasks. These points complement recent work around explanations for AI-enabled and algorithmic systems, where explanations likewise are argued to be ``inherently contextual'', relating to both the target group and task addressed by an explanation.\footnote{\cite{asghari2021explain}. See also \cite{krishnan2019against}.} Explanations and definitions share many core elements, and while one can argue for ``the'' definition of some term, object, or entity, it is far more fruitful to allow for numerous definitions tailored to target groups and tasks. 

% We are engineers first, scientists second
	% The aim is to provide something useful, even if it is not necessarily 100% ``true''
% Our definitions ought to be scientifically defensible, but must also be broadly acceptable to non-scientific audiences/stakeholders
If we aim to develop a base definition which can be refined and specified along broadly pragmatic grounds as needed, with the usefulness of that definition being a central criterion, a number of considerations must be taken into account. First, the base definition should be scientifically defensible while also understandable and acceptable to non-scientific audiences and stakeholders. Without this, it may be technically accurate, but unhelpful in application. 

% The job of our definitions is to provide a starting point for exploration, not to foreground conclusions we hope to draw
% Definitions should be neutral; i.e., not prejudicial to follow-on debates which make use of that definition
	% A view of ``autonomy' including in situ or ``online'' learning is prejudicial to the debates
	% In situ learning would make systems almost certain to be in violation of International Humanitarian Law (IHL) 
		% IHL excursus: Additional Protocol I, Art. 36: ``In the study, development, acquisition or adoption of a new weapon, means or method of warfare, a High Contracting Party is under an obligation to determine whether its employment would, in some or all circumstances, be prohibited by this Protocol or by any other rule of international law applicable to the High Contracting Party.''
% Any view of autonomy that makes it impossible to control, understand, or predict how a system will behave should be deemed prejudicial
It should serve as a starting point for further discussion and debate, and not foreground or anticipate later conclusions we may hope to draw. Building on this, it should be neutral in three distinct respects. First, it should be non-prejudicial, meaning that the definition itself ought not unduly color or prejudge follow-on debates about the thing being defined. For example, in the debates around autonomous weapon systems, a view of ``autonomy'' which includes \textit{in situ} or online learning as a necessary definitional component confers by fiat a host of arguably insurmountable challenges for AWS. This is because systems which can learn in real time and potentially alter their program almost certainly cannot be subjected to the necessary test, evaluation, verification, and validation required for weapon systems to be permissibly fielded in military contexts at all.\footnote{See, e.g., \cite{blanchard2022predictability,haugh2018status,verbruggen2022no,macfarland2023legal,zajac2026inexplainability}. For general discussion of the legal requirements for weapons review, see, e.g., \cite{boothby2016weapons,jevglevskaja2018weapons} and for review challenges specific to emerging technologies and autonomous weapons, see \cite{boothby2014legal,jevglevskaja2021challenges,jevglevskaja2021international}.\label{ftnote_in-situ}} Thus, any view of autonomy that makes it impossible to control, adequately understand, or sufficiently predict how a system will behave should be deemed prejudicial to knock-on discussions. In some instances, the ``right'' definition may by necessity color further debate, but when there are reasonable alternative definitions which do not include a prejudicial framing, these are to be preferred.\footnote{See \cite{corlett1996can,schwenkenbecher2012terrorism} for discussion of the need for neutral definitions of terms.}

% A neutral definition should also be time-neutral (i.e., not aim to ensure that ``autonomous systems'' do not yet exist)
Base definitions should also ideally be time-neutral, meaning that they are not designed to ensure that the defined systems (do not yet) exist. In some discussions of AWS, authors are keen to point out that the thing they are defining doesn't exist yet (though they are often incorrect in their presentation; see \cite{wood2023awsclarification} and citations therein). In other definitions, such as that until recently provided by the United Kingdom's Ministry of Defence (UK MoD), a highly strict definition has been given which allows some to, under that view, claim they do not possess autonomous weapons and do not plan to explore these.\footnote{Precisely, the UK MoD in \textit{Joint Doctrine Publication 0-30.2: Unmanned Aircraft Systems} defined ``autonomous systems'' as those ``capable of understanding higher-level intent and direction'' (\cite{ukmod2017unmanned}, p. 13), before stressing that ``[t]he UK does not possess fully autonomous weapon systems and has no intention of developing them'' (\cite{ukmod2017unmanned}, p. 14). This earlier document has been superseded by \textit{Joint Doctrine Publication 0-30: UK Air Power}, which softens the strictness of the UK's definition, though not to as broad a view as that presented by the ``standard'' view, and reiterates that ``[t]he UK does not possess armed autonomous aircraft systems, and has no intention of developing them'' (\cite{ukmod2022air}, p. 12).} That strict definition appears to be designed to satisfy a political aim (proclaiming their rejection of AWS, as defined by them), but does so at the cost of defining the term in a highly unhelpful manner (in virtue of the definition applying to a highly narrow class of entities, and one out of step with the current broad understanding).\footnote{See \cite{taddeo2022comparative}, pp. 9--10 for further exploration of the UK MoD definition, and related statements from other states.} Building definitions to satisfy an underlying commitment about whether or not some thing exists represents a type of prejudice harmful to effective discussion. 

% A neutral definition should be use-neutral (i.e., aim to allow for definitions of systems which can be used in various ways, for various ends, with various levels or permissibility)
The third kind of neutrality in need of incorporation concerns use. Base definitions should be use-neutral, meaning they do not tie a broad class of technology to a narrow domain of application. I.e., one and the same base definition for ``autonomous weapon systems'' should be able to capture AWS used in anti-personnel or anti-materiel roles, with lethal or less-lethal armaments on board, used in purely defensive or potentially defensive/offensive platforms, etc. Refinements on that base definition can pick out more specific platforms -- Section \ref{sec_conceptual-framework} explores a variety of such refining categories -- but the basic definition of AWS should not prejudge what more specific configurations of systems and capabilities fit within that. This is especially the case, given that some capabilities or use cases may be more ethically, legally, or operationally fraught than others, and so tying the base definition to one case (e.g., anti-personnel systems) will prejudice all following discussions. 

% Recap: a definition should be 1) as accurate as possible while being 2) useful, 3) acceptable to scientists and lay stakeholder groups, 4) non-prejudicial, 5) time-neutral, and 6) use neutral
To briefly recap, a base definition should be 1) as accurate as possible 2) while still being broadly useful, 3) acceptable to scientists and non-scientific stakeholder groups alike, 4) non-prejudicial in general terms, 5) time-neutral regarding whether the defined object exists (or must exist for the defining party's aims), and 6) use neutral regarding whether the system must by necessity execute certain tasks or have certain target sets. 

\section{Existing Views and Their Limitations}\label{sec_existing-views}

With these considerations and desiderata for a ``good'' base definition in place, we may now critically evaluate two broad views of AWS proposed. The first we may think of as the ``standard'' view, shared by key state actors, international organizations, and many scholars. The second, which we call the ``demanding'' view, presents a much narrower view of AWS, and while not explicitly endorsed by as many groups or individuals, finds implicit voice in many arguments critical of AWS. The main aim of the criticisms leveled in this section is to draw out errors and points relevant for the creation of a better base definition upon which refinements may be layered, according to particular more specific systems or definitional use cases. 

\subsection{The Standard View}\label{subsec_existing-views_standard}

What we have termed the ``standard'' view is that prominently presented by the United States Department of Defense (US DoD), the International Committee of the Red Cross (ICRC), Human Rights Watch, and numerous other states, international organizations, non-governmental organizations, and scholars. This is the conception of AWS as weapon systems with autonomy in ``critical functions'' necessary for selecting and engaging targets,\footnote{\cite{icrc2014autonomous}, p. 5.} and which can therefore select and engage targets without contemporaneous human input, oversight, or intervention.\footnote{\cite{icrc2021positiona}, p. 1; \cite{usdod2023directive3000.09}, p. 21. See also \cite{williams2015defining,boothby2016weapons,altmann2017autonomous,caron2020defining,taddeo2022comparative} for overviews and taxonomical work; \cite{pacholska2024autonomous} for comparative discussion of core definitions from central actors.} As noted by \cite{wood2024regulating}, the ``broad view is similarly shared by both scholars critical of AWS... and civil society groups'',\footnote{\cite{wood2024regulating}, p. 2, referencing \cite{asaro2012banning}, p. 690; \cite{roff2014strategic}, p. 353; \cite{chengeta2016measuring}, p. 67; \cite{sparrow2016robots}, p. 95; \cite{asaro2019algorithms}, p. 540; \cite{hrw2018heed}, p. 6; \cite{hrw2020new}, p. 2; \cite{skr2022negotiating}, p. 6.} highlighting that this definition is not merely the preferred view for states and regulatory bodies, but sees broad acceptance across various domains. As an exemplar of the standard view, we may take the definition given by the US DoD:

\begin{quote}
	A weapon system that, once activated, can select and engage targets without further intervention by an operator. This includes, but is not limited to, operator-supervised autonomous weapon systems that are designed to allow operators to override operation of the weapon system, but can select and engage targets without further operator input after activation.\footnote{\cite{usdod2023directive3000.09}, p. 21. The recent meetings of the Group of Governmental Experts (GGE) at the CCW also resulted in an agreed definition, but of \textit{lethal} autonomous weapon systems (LAWS). That definition largely tracks that presented here from the US DoD, with the addition that LAWS are ``capable of identifying, selecting, and engaging a target \textit{in a way that results in loss of life}, without intervention by a human being operating the system'' (\cite{gge2026reportrev3}, p. 4, emphasis added). In line with the arguments to come, this definition from the GGE is also far too broad though, as many systems designed to engage materiel exclusively are ``capable of'' engaging targets ``in a way that results in loss of life'', for example, when debris from engaged targets falls on inhabited areas. For documentation of the GGE's proceedings, see \cite{ccwgge2026ngoreport}.}
\end{quote}

The standard view anchors a central aspect of AWS, that they are autonomous in core tasks relevant to a weapon system's contribution to military operations. Beyond that, this view does not imply any other specific capabilities, limitations, or design choices though, allowing for AWS which are purely mechanical or which have the most sophisticated AI-enabled functions; which are transparent or opaque in their functioning; which are designed to be lethal or less lethal; or which exclusively target persons or objects. Given this breadth, there are many classes of system which have existed for decades which count as AWS under this definition, from anti-radiation missiles to missile interception systems, and even landmines.\footnote{\cite{sipri2020limits,icrc2021positionb,heller2023concept,wood2023awsclarification,wood2023awsresponsibility}.} This presents a basic challenge on its own, and there are three specific implications of the broad definition which undermine its usefulness for debates. 

% Includes biological entities (e.g., bat bombs); taking a system of systems approach to warfighting, the human combatant can also be viewed as an ``autonomous weapon system''
First, the debates around AWS and emerging technologies in the military more broadly are concerned with \textit{artifacts} humans make and use. However, there are ``weapon systems'' which are autonomous and based centrally around biological entities. From the ``bat bombs'' of World War II\footnote{During World War II, research was conducted under Project X-Ray into the viability of attaching incendiary devices to bats as a means of delivering ordnance to the Japanese heartland. See \cite{couffer1992bat,crane2008bats}.} to modern combat-assault-dogs,\footnote{\cite{waller1958dogs,hammerstrom2005ground,lemish2008war,rogak2011dogs}.} and many other animals besides have been used in war throughout the ages.\footnote{\cite{nowrot2015animals,briefer2024goats}.} Taking for granted the system-of-systems understanding of modern military operations, where overall military potential is viewed as a result of the ``interactions and synergism''\footnote{\cite{manthorpe1996emerging}, p. 309. See also \cite{owens1996emerging,chen2017sos}.} of distinct systems, each contributing to a larger aim, it is also possible to view individual human combatants as ``autonomous weapon systems'', given that they are capable of selecting and engaging targets without further input and are systems within the larger system-of-systems driving modern warfare.\footnote{Less significantly but in a similar vein, some views define human personnel operating a system as part of the definition of a ``weapon system''. E.g., the NATO glossary agreed definition of ``weapon system'' is ``A combination of one or more weapons with all related equipment, materials, services, personnel and means of delivery and deployment, if applicable, required for self-sufficiency'' (NATO Glossary Record 777, findable through \url{https://nso.nato.int/natoterm}). Thus, even when humans are not to be considered AWS themselves, they may under existing definitions and doctrine be deemed a necessary part of autonomous weapon systems.} One could bite the bullet and claim that ``autonomous weapon systems'' can reasonably include biological entities, humans included, within its scope, but this would render the term highly unfit for the regulatory and doctrinal roles it is meant to fulfill.

% Assumes functionality; a broken AWS may fail the test of whether it ``can'' select and engage targets, but that should not render it not an AWS
A second concern is that the standard view's focus on systems that \textit{can} select and engage targets assumes a certain level of functionality in order for objects to count as AWS. Put differently, it excludes systems that might be broken or improperly functioning. However, if systems are designed to ``select and engage targets without further intervention'' and normally can do that, temporary malfunctioning should not remove them from the definitional extension of AWS. This may seem a small quibble, but definitions set the standard for not just discussions, but also legal regulations, and a regulation governing AWS which assumes functionality will, on a strict reading, not cover any systems malfunctioning or temporarily limited. This is a problem. 

% Is generally too broad; there are so many things falling under this definition, potentially even including things that fall outside the traditional idea of warfighting as a centrally kinetic enterprise (e.g., chatbots conducting autonomous information operations may be AWS, and thus captured without notice under the same heading as a term used for roving missile-armed drones)
Finally, the standard view is, quite simply, too broad in general terms. There are very many weapon systems which fall under this definition, used for an exceedingly wide array of tasks, and with widely disparate capabilities, use contexts, ethical and legal challenges, and other attendant concerns. Focusing on the mere capacity to ``select and engage targets without further intervention by an operator'' also implies that things which plausibly fall outside the traditional idea of warfighting would also count as AWS. For example, chatbots fine-tuned to carry out information operations and given agentic AI scaffolding to allow them to independently engage with foreign civilians online could also be seen as AWS, simply because they could select targets (online users) and engage them (post messages tailored to them) without further input. In the evolving landscape of modern conflict, greater attention should be paid to the information domain (see Section \ref{subsec_conceptual-framework_io-cognitive} below), but capturing a missile-armed autonomous drone and a chatbot posting to social media under the same heading is clearly not helpful. 

% This definition is very broad; for some applications/uses, that is useful (i.e., for highest-level regulations or norms), but that breadth means that much of the time the definition will not be very helpful; having landmines and autonomous AI-enabled attack drones that can operate over very long distances and open timeframes in the same definition is fairly ridiculous
To sum, the standard view is very broad. For some applications, such as establishing fundamental norms or discussing elementary concerns over control and oversight, this breadth may be useful. But in the vast majority of use cases, the standard view will be too broad, covering such a wide array of systems that it loses its value and serves more to muddy discussion than to further it. For these reasons, a more precise view is worth pursuing. 

\subsection{The Demanding View}\label{subsec_existing-views_strict}

There are a variety of ways the standard view may be given more precision, or its extension limited. Across the literature, a standard approach is to simply add constraints or definitional points which make the definition of AWS more demanding. Above, we have already seen this with the UK MoD definition (now superseded) of AWS as systems ``capable of understanding higher-level intent and direction'',\footnote{\cite{ukmod2017unmanned}, p. 13.} and the Canadian discussion of AWS admits similar addition of terms indicative of higher agency in AWS.\footnote{\cite{shapiro2019autonomous}, p. 2, and citations therein.} While there are numerous ways to develop a more demanding view, for the remainder of this section we will focus on the definition given by \cite{taddeo2022comparative}, as it represents one of the most fully developed, motivated, and defended instances of the ``demanding'' view. They define an autonomous weapon system as follows:

\begin{quote}
	[A]n artificial agent which, at the very minimum, is able to change its own internal states to achieve a given goal, or set of goals, within its dynamic operating environment and without the direct intervention of another agent and may also be endowed with some abilities for changing its own transition rules without the intervention of another agent, and which is deployed with the purpose of exerting kinetic force against a physical entity (whether an object or a human being) and to this end is able to identify, select or attack the target without the intervention of another agent is an AWS. Once deployed, AWS can be operated with or without some forms of human control (in, on or out the loop). A lethal AWS is specific subset of an AWS with the goal of exerting kinetic force against human beings.\footnote{\cite{taddeo2022comparative}, p. 15.}% N.b., Taddeo and Blanchard use ``AWS'' as an abbreviation for ``autonomous weapon\textit{s} system\textit{s}” (i.e., the plural form) which should be avoided, as the aim of a definition should be to allow us to identify singular artifacts, not merely a collection of AWSs.}
\end{quote}

% Generally too long, which undermines some basic pragmatic concerns in definitional construction
% Pros coming from added specificity; autonomous state-changing to meet operational needs; autonomous program changing to meet operational needs; used to exert kinetic force against physical entities; can be deployed with or without human control; can be ``lethal'', meaning explicitly anti-personnel
Taddeo and Blanchard's definition is unfortunately lengthy and complex, undermining its potential use in many domains where simplicity is a virtue. Even so, it gives much added specificity to the core conception of ``autonomous weapon systems''. Inclusion of autonomous changing of internal states and transition rules are far more demanding standards than those present in the standard view, greatly limiting the extension of AWS. The stress on kinetic force also eliminates the possibility for systems usually wholly outside direct military operations -- such as autonomous information warfare systems -- to be considered AWS. Finally, addition of varying levels of control, lethality, and target class (personnel vs. materiel) as potential modifiers serve as useful ways of framing nuances in system design and deployment. Yet for all its added specificity, the demanding view presented by Taddeo and Blanchard critically fails for a number of reasons, with the failures indicating general issues apt to obtain for all such demanding views. 

Before examining the individual distinct shortcomings of Taddeo and Blanchard's definition, it is worth clarifying that in this section, we will not be evaluating their definition against their own aims\footnote{Specifically, they claim to, ``offer a value-neutral definition of AWS. In doing so we have a twin-goal of (i) defining the key characteristics that permit the identification of AWS; and (ii) specifying these characteristics so to clarify their relations--e.g. automation vs control, and their differences--e.g. automatic vs autonomous'' (\cite{taddeo2022comparative}, p. 14).} (though their definition does fail on some of their own aims; see discussion of value-neutrality below). Rather, we will focus on whether the extension of their definition sensibly tracks to what is broadly considered the ``correct'' extension of AWS. As artifacts do not fit as cleanly into taxonomic classes as natural and scientifically discernible entities do, some reference to general usage is an apt limiter of the success or failure of a definition. Moreover, whether the extension of a definition fits general usage is also critical to whether that definition is useful in terms of furthering, guiding, or grounding ongoing debates. If the extension of a definition departs widely from standard usage, then it is not a useful definition, and may indeed be a harmful one, in virtue of adding confusion to an already quite complex and multi-disciplinary enterprise with significant human consequences attending misunderstandings. Given these points, we will primarily critique the demanding view presented by Taddeo and Blanchard in light of how well it fits to the broad consensus in international debates concerning what are (not) AWS.

% Cons: 
% Artificial agent is too broad and narrow; to broadness, a clone would arguably be an ``artificial'' agent; clone would properly be treated as a fully-fledged combatant, not \textit{merely} an AWS; even under the system-of-systems view, where individual combatants may be viewed as subsystems in the larger military system, it is still awkward at the least, and likely simply wrong to view human combatants as ``autonomous weapon systems'', as this greatly misrepresents their agency, ethical and legal responsibility, and capacity for novel action; to narrowness, an artificial ``agent'' sets the bar for AWS too high, as agency can -- and for ethical and legal reasons, should -- be viewed as a thick concept which would naturally rule out any but the most thoroughly advanced science fiction machine systems as ``agents''; requiring that AWS be agents thus relegates the definition to far future science fiction, making it a useless conception for ongoing debates
Problems arise already with the very first terms in Taddeo and Blanchard's definition, as ``artificial agent'' renders the definition both too broad and too narrow. To broadness, human clones would arguably be ``artificial'' agents, given that they are artificially created beings. However, to treat clone as \textit{merely} AWS would be technically, morally, and legally incorrect. Whether a human is the result of natural reproduction or some futuristic cloning process, such a human is still a human and therefore a full moral and legal agent, and should be treated as a fully-fledged combatant, with all the rights and responsibilities accorded that status. Even under the system-of-systems view presented above, where individual combatants may be viewed as subsystems in the larger military system, it is still awkward at the least, and likely simply wrong to view human combatants as ``autonomous weapon systems'', as this greatly misrepresents their agency, ethical and legal responsibility, and capacity for novel action. To narrowness, artificial ``agents'' sets the bar for AWS too high, as agency can -- and for ethical and legal reasons, should -- be viewed as a thick concept that naturally rules out any but the most thoroughly advanced science fiction machine systems as ``agents''. This is because agency implies the ability to take responsibility and to be reasonably held responsible for genuine choices one makes.\footnote{\cite{ginet1997freedom,kennett2003agency,moor2006nature}.} This is what distinguishes in a court of law humans of normal capacity from those who are (temporarily) insane or who are not fully cognitively capable (children or the mentally handicapped).\footnote{\cite{traina2009children,tiboris2014blaming}.} That distinction matters in ethics as well, and is a critical factor in the organization of military command structures and responsibility chains. Requiring that AWS be agents thus relegates the definition's extension to far future science fiction, making it a useless conception for ongoing debates.

% Assumption of functionality is also present, making the definition too narrow; reiterate briefly points from section above
The demanding view presented by Taddeo and Blanchard also admits the same assumption of functionality found in the standard view, in light of picking out systems ``able to change'' internal states and ``able to identify, select or attack'' targets. As discussed above, this renders the definition too narrow, as systems designed to be autonomous but temporarily malfunctioning should not be excluded due to passing system limitations. The focus must be on design architectures and intended functionalities, not the precise capability set possessed by a system at a time of evaluation. 

% Focus on kinetic force makes the view too narrow; there are very serious non-kinetic forms of attack that may be carried out, and may be carried out by autonomous systems, and excluding a non-kinetic autonomous weapon system from the class of AWS simply because it does not ``go boom'' is flatly incorrect; mention EW, cyber, or info/cognitive as various realms where non-kinetic action can be quite significant
Focusing on systems ``exerting kinetic force'' is likewise too narrow, as modern warfare includes a variety of highly impactful non-kinetic forms of attack that may be carried out by autonomous systems. For example, electronic warfare (EW), cyber-attacks, or highly targeted information/cognitive warfare can be quite significant to ongoing military operations, with EW and cyber-attacks being central elements in modern drone-intensive and sensor-linked warfare. Limiting the extension of AWS to only those things that have immediate kinetic effects is thus unduly narrowing in general terms, and also out of touch with central realities of modern warfare and emerging autonomous systems design.\footnote{See, e.g., \cite{lehto2020non}.} 

% Focus on ``physical entities'' as the sole target class is also overly narrowing; digital infrastructures have a physical side to them, but autonomous cyber attacks may be aimed solely at the digital element (noting that every digital element has a corresponding physicality to it); some attacks are not against systems themselves either, but are aimed at ``capacities'' in systems, which is hard to see as a physical thing (e.g., jamming and spoofing attacks aim to undermine a systems ability to navigate or communicate well, but do not necessarily target the platform itself; indeed, many such attacks are area attacks, and so are not aimed at any one platform, but rather serve as capability-denial attacks)
Related to this, picking out ``physical entities'' as the sole target class of AWS is also unduly narrow. Modern warfare increasingly relies on digital infrastructures to underpin core military capabilities,\footnote{\cite{bray2026hardware}.} and while it is true that digital infrastructures have a physical side to them, there are potentially autonomous attacks that may be aimed solely at the digital aspects of such systems.\footnote{We note that no system is ``purely digital'', as every line of code and bit of compute has a physical analog and embodiment somewhere. That being said, there are cyber-attacks that target the logic or structure of a code, i.e., its immaterial element, and do not primarily aim to disrupt the physical instantiation of that code or compute. Other attacks may physically target a server bank, data center, internet hub, or other physical embodiment of digital systems.} There are also attacks aimed not at systems themselves \textit{per se}, but rather at capacities in systems. For example, jamming and spoofing attacks\footnote{\cite{vakin2001fundamentals,zhang2023radar}.} aim to undermine a system's ability to navigate or communicate, but do not necessarily target the platform itself. Indeed, many such attacks are area attacks, and so are not aimed at any one platform, but rather serve as capability-denial area attacks. It is hard to see how an electronic attack intended to undermine a capacity to navigate represents a strike on a ``physical entity''. 

% Limit to ``purpose of'' conflates what ``is a weapon'' with what ``can be used as a weapon'', making the definition too narrow and potentially too broad; too narrow because under this definition, autonomous weapon systems would cease to be AWS any time they were deployed without the intent to exert force (e.g., deploying an autonomous Reaper drone for ISR operations would render it not an AWS, but potentially only for that deployment); it could be too broad as well, as it would capture autonomous non-weapons as AWS whenever they are used for attacks (e.g., using an autonomous car and having it drive through the gates of an enemy outpost); perhaps such a use-determinant definition could have value, but it seems more natural to say in such examples that ``this is an autonomous car \textit{being used as an AWS}, but not that it \textit{is an AWS}''; we leave this point somewhat open, as there are good arguments to be made for both sides, but we point out that the clarification just provided is more parsimonious than having the extension of the term change according to momentary use; we also stress that while ``it is used an an AWS'' and ``it is an AWS'' may seem a trivial and pedantic point to make, such pedantry is critical for legal regulations, lest norms governing weapons bleed over into purely civilian domains, or fail to govern the objects they are meant to oversee
Limiting the definition to systems ``deployed with the purpose of'' exerting force also conflates what \textit{is a weapon} with what \textit{can be used as a weapon}, making the definition too narrow and potentially too broad as well. It is too narrow because this restriction implies that autonomous weapon systems cease to be AWS any time they are deployed without an intent to exert force. For example, deploying missile-armed autonomous aerial drones for purely reconnaissance purposes would render them not AWS. Oddly, they would also potentially not be AWS \textit{only for that operation}. This is clearly incorrect, as the purpose of a deployment should not alter the essential definition of the systems being deployed, especially considering that the purpose may change mid-deployment (say ``reconnaissance-only'' missile-armed drones identify enemy positions, and then are switched by human overseers into ``reconnaissance and strike'' mode). 

Focus on the purpose of a deployment may be too broad as well, as it implies that some objects not normally considered weapons may be identified as AWS whenever they are used for attacks. For example, autonomous cars are not AWS, but if an insurgent were to modify/remove crash-avoidance systems and then use an autonomous car as a battering ram to strike the gates of a military base, it would, under this definition, be an autonomous weapon system \textit{for that operation}. Perhaps such a use-determinant definition can have value, but it seems far more natural to say in such cases that ``this is an autonomous car \textit{being used as an autonomous weapon system}'', not simply that it ``\textit{is an autonomous weapon system}''. To be sure, there may be good arguments for both sides of this particular point, but the clarification just provided is much more parsimonious than having the extension of a term change according to momentary use. We would also stress that while ``they are used as AWS'' and ``they are AWS'' may seem a trivial and pedantic point to make, such pedantry is critical for legal regulations, lest norms governing weapons bleed over into purely civilian domains, or fail to govern the objects they are meant to oversee. In short, the precise extension of a definition can unduly limit or unfairly expand legal regimes, and so attention to that extension is merited. 

% This is quite simply not value-neutral (violating their own avowed aims); any system that can autonomously change not only its states but also its transition rules is inherently utterly unpredictable, and thus is by definition an indiscriminate system (as it cannot be adequately directed and its effects cannot be limited, AP I, Art. 51.4.b-c)
The final and arguably most poignant concern with Taddeo and Blanchard's view is that it simply is not neutral, in violation of both a central desiderata of good definitions discussed in Section \ref{sec_definitions} above and Taddeo and Blanchard's own avowed aim to ``identify essential aspects of AWS and then offer a definition that provides a \textit{value-neutral ground} to facilitate efforts to address the relevant ethical and legal problems''.\footnote{\cite{taddeo2022comparative}, p. 5, emphasis added.} This is because, as discussed above, merely including \textit{in situ} or online learning in the definition of AWS would render these systems incapable of being used in compliance with core rules of international law concerning necessary test and evaluation of new systems.\footnote{See note \ref{ftnote_in-situ} and references therein.} The demanding definition above goes far beyond \textit{in situ} learning though, anchoring an ability to autonomously change internal states and potentially also programming as essential features of AWS. Quite bluntly, it would be fundamentally illegal to deploy any system which had such a capability, both because it could not be ensured that its use would not, ``in some or all circumstances, be prohibited by this Protocol or by any other rule of international law'', in violation of Article 36 of Protocol I Additional to the Geneva Conventions (hereafter, AP I),\footnote{AP I, Art. 36.} and because the potential unpredictability of such a system would render it a ``method or means of combat which cannot be directed at a specific military objective'' or a ``method or means of combat the effects of which cannot be limited'', violating Article 51 of AP I.\footnote{Specifically, AP I, Art. 51.4.b-c. For extensive discussion of distinction, AWS, and the realities of modern warfare see e.g., \cite{schmitt2012autonomous,zajac2022autonomous,heller2023concept,pacholska2023military,wood2026bombs}.} A definition of AWS which renders their use necessarily incompatible with international humanitarian law is not a neutral definition.\footnote{Note also that Taddeo and Blanchard in immediately subsequent publications rely on this definition to ground core critiques of AWS, critiques that require the assumption of \textit{in situ} learning and continuous autonomous updating of systems. See, e.g., \cite{blanchard2022predictability,blanchard2023jus}.} Moreover, it is not a definition which facilitates debate or efforts to address ethical and legal concerns, as the only reasonable conclusion to draw from Taddeo and Blanchard's view is that AWS, as understood by them, must be explicitly and comprehensively banned, as they could not be used in compliance with international humanitarian law. 

% Demanding views in general will have to admit additional terms that add the stringency necessary for limiting the definition's extension; every such addition is going to remove some (class of) systems which plausibly fit under the umbrella of ``autonomous weapon systems''; rather than trying to have a single unary definition that captures all plausible candidates of AWS (which is important for regulatory, oversight, and doctrinal reasons), we should aim for a base definition that captures essential features grounding broad-based fundamental norms, and then have a modular approach to constructing refined definitions that are system- and use-specific
The demanding view developed by Taddeo and Blanchard suffers from many faults specific to their particular formulation. Broadly though, demanding views will in general need to admit additional terms or definitional requirements in order to add stringency for limiting those definitions' extensions. Every such addition will remove some (class of) systems which plausibly fit under the umbrella of ``autonomous weapon systems'' for some reasons or with respect to some use cases or purposes. Rather than trying to develop a singular unary definition that captures all plausible candidates of AWS (which can be important for regulatory, oversight, or doctrinal reasons), we should aim for a base definition which captures the essential features of AWS grounding broad-based fundamental norms, and then have a modular approach to constructing refined definitions that are system-, use-, and purpose-specific. 

\section{The Base Definition}\label{sec_def.-aws}

%The standard view of AWS is too broad and gives little guidance for how to systematically parse or group sub-concepts captured under ``autonomous weapon systems''. The demanding view is very likely to exclude important (classes of) systems, and the means of restricting the definition's extension are also apt to raise problems of their own (as argued above). As such, we should aim for a broad but sensible base definition capable of grounding fundamental debates and norms, and then establish a catalog of modular elements that can be plugged into refined definitions to make them system- and use-specific.
A base definition of ``autonomous weapon systems'' must be broad enough to ground fundamental ethical, legal, doctrinal, and governance norms which span various systems and use cases. It must also not prejudge systems. And finally, it must be loose enough in its definitional stringency and structurally permissive in allowing other terms to be added into it as needed.  

The standard view presented above meets most of these criteria, and while being too broad to be acceptable as \textit{the} definition of AWS, it could serve well as a base definition, upon which additional elements are grafted to suit particular needs. However, as argued in Section \ref{subsec_existing-views_standard}, there are two specific elements of the standard view in need of addressing, namely that it includes biological entities as AWS, even potentially human combatants, and that it assumes that systems classified as AWS are properly functioning. Each of these must be addressed in a working base definition. 

% Autonomous System: A machine system that, if properly functioning, once activated, can independently identify and pursue (sets of) processes which further a broader goal, task, or mission set by a human deployer.
Given the ultimate aim of developing a modular framework, there is also value in beginning not with a base definition of ``autonomous weapon systems'', but instead with the more general ``autonomous systems''.\footnote{See \cite{sillitto2019systems} for additional basic definitions of elementary terms.} The ability to engage in warfighting functions -- i.e., the term ``weapons'' -- then represents the first basic module to be added. The base definition of ``autonomous systems'' may be augmented with non-weapon elements to serve in more specific definitions in other domains as well, or in non-combat aspects of the military such as bureaucratic, logistical, or other support roles.\footnote{For discussion of various uses of AI-enabled systems in the military, see \cite{wood2026stop}.} This most basic definition may be given as follows:

\begin{quote}
	\textbf{Autonomous Systems (base definition):} Machine-based systems that, if properly functioning, once activated, can identify and pursue (sets of) processes without contemporaneous human input which further a broader goal, task, or mission set by a human deployer.
\end{quote}

By picking out a ``machine-based system'', purely organic entities are excluded from the class of autonomous systems, and the assumption of proper functioning in this base definition removes the risk that malfunctioning systems are temporarily excluded from its extension. The selection of ``machine-based'' rather than simply ``machine'' also allows for inclusion of bio-robotic elements in systems without this automatically removing these from the base definition's scope. %\footnote{For a review of recent advances in bio-robotics and their potential implications and applications, see CITATION.} 
Beyond these modifications, the base definition roughly follows the intent and framing of the standard view presented by the US DoD and the ICRC, but with respect to autonomous systems as such. Thus, the term tracks to autonomous execution of processes more generally, but not necessarily such processes with an explicitly warfighting function to them. For such a warfighting-centric base definition of ``autonomous weapon systems'', we give the following:

% Autonomous Weapon System: A machine weapon system that, if properly functioning, once activated, can identify, select, and engage targets without contemporaneous human input. (ICRC/US DoD)
\begin{quote}
	\textbf{Autonomous Weapon Systems (base definition):} Machine-based weapon systems that, if properly functioning, once activated, can identify, select, and engage targets without contemporaneous human input.
\end{quote}

As with the previous base definition, ``machine-based'' and ``if properly functioning'' serve to rule out those problem cases of the standard view discussed in \ref{subsec_existing-views_standard} above. By adding identification, selection, and \textit{engagement} of targets, the base definition of ``autonomous systems'' can be made more precise for the warfighting aspect of the military domain. 

The reader might here object that the proposed base definition's similarity to the standard view risks inheriting its weaknesses. In response, we stress that the core problem of the standard view was not that it was ``wrong'' \textit{per se}, but rather that it was simply too broad.\footnote{Though there are also extensional problems of the standard view which our proposed base definition addresses, such as the concerns around functioning AWS or biological AWS.} The base definition thus takes the clarity and simplicity of the standard view, two virtues in a definition, and avoids the problems of over-broadness by anchoring that this is a \textit{base} definition only. Thus, it only picks out the most basic and essential elements of an AWS, but does not articulate all or even necessarily the most important elements of some system for a given use-case or oversight aim. To achieve those more specific aims, the base definition must be augmented by modular terms picking out core concerns that may motivate ethical, legal, doctrinal, technological, or oversight debates. Yet even without those additions, central failings of the US DoD/ICRC standard view can be avoided, and the base definition serves as a useful grounding for discussion of most-fundamental regulations and oversight principles. For anything beyond those, added specificity will, however, be necessary. 

\section{A Conceptual Framework for Refined Definitions}\label{sec_conceptual-framework}

% Start with system; build ``weapon'' onto that definition; then iterate off of that with other details; clarify that without this additional iteration, the base definition is not very useful for many purposes (except perhaps the development of minimal doctrine, standards, rules, laws, etc.; reference \cite{wood2026stop} for this)
Starting with a base definition of ``autonomous systems'', ``autonomous weapon systems'' may be easily defined as a subcategory endowed with particular capabilities and fulfilling a particular role. The base definition of ``autonomous weapon systems'' given above is still too broad to be of use outside of most-fundamental regulatory, technical, or doctrinal discussions though. For any debates focusing on systems with specific levels of autonomy, used for particular combat functions, having additional capabilities, or otherwise drawing on more specific particularities of design and use, the base definition must be augmented with additional terms. The central aim of this section is to present a variety of such terms, focusing first on four elements which are apt to be central to core debates around AWS, after which we briefly canvas six narrower elements relevant for specific platforms or warfighting domains, and finally giving a brief guide to designing definitions for particular use cases. Having a catalog of such terms and a guide for how to leverage them for refined definitions is also critical, as without these, the base definition will not be useful for any purposes outside of establishing the most basic fundamental norms for AWS.\footnote{See \cite{wood2026stop} for similar points.} 

% We will list many modules which make sense or are natural candidates for refining the base definition as needed for a specific system and use-case; we will however certainly not catch all important details; we also note that we will omit discussion of systems which \textit{can be used} as AWS, focusing instead on systems \textit{designed to be} AWS in the course of their normal functioning (See Section \ref{sec_existing-views}, and also \cite{franssen2024technology} for discussion of the distinction ``between the \textit{proper function} of an artifact as the one intended by its designer and the \textit{accidental function} of the artifact as the one given to it by some user on private considerations'')
It should be noted that while we present many terms which are natural candidates for refining the base definition in light of specific systems and use-cases, we do not claim our presentation to be complete. Thus, there may be other terms not captured by our discussion which can be included in a modular fashion to augment the base definition.\footnote{The discussion of \cite{williams2015defining} covers many elementary terms which could plausibly have value for different definitional use cases. That discussion serves as a vital departure point for much of our arguments. See also Ch. 14 of \cite{boothby2016weapons}.} This is a central boon of having a \textit{modular conceptual framework} for AWS rather a single unary definition treated as ``the conception''. Given that emerging technologies are constantly changing, that warfare is driving rapid innovation, and that militaries are finding new ways of using existing technologies, it is preferable to have a framework for definitions rather than a single definition which is ill-suited to many use cases. 

We also note that for reasons of space we will omit all discussion of systems which \textit{can be used} as AWS, focusing instead on only those systems \textit{designed to be} AWS in the course of their normal functioning.\footnote{See Section \ref{sec_existing-views} and also \cite{franssen2024technology} for discussion of the distinction ``between the \textit{proper function} of an artifact as the one intended by its designer and the \textit{accidental function} of the artifact as the one given to it by some user on private considerations''.} Systems which can be used as AWS may also be captured under our definitional schema though by simply taking the base definition of ``autonomous system'' and adding to it that such systems are used -- but not necessarily designed -- to engage or attack some persons, objects, or capabilities. The modular schematic construction thus allows one to examine more than just AWS, and lays out refined definitions in a clear and traceable way, with all relevant points easily demarcated and able to be interrogated as needed.

\subsection{Autonomy, State Transitions, and Self-Learning}\label{subsec_conceptual-framework_autonomy}

% This section focuses on levels of autonomy in a given system; despite presenting a definition which is untenable due to its prejudicial nature, Taddeo and Blanchard's discussion of autonomy, state transitions, and self-learning is quite helpful (see esp. pp. 15--17)
One of the first and arguably most important elements to add to a base definition is what precise level of autonomy a system has, what internal states it can change on its own, and whether it has the capacity to add to, rewrite, or otherwise modify its own programming, approaches, or goals. Autonomy is moreover a critical term to have explicitly defined in a more refined definition of AWS (or autonomous systems generally) given that the term has distinctly different understandings in the philosophical/ethical/legal traditions compared to how it is used in engineering/computer science research. In the former, autonomy is deeply entwined with being a ``first mover'' or free agent that can make genuine decisions in the world, whereas the latter view focuses simply on independent execution of tasks without direct oversight or intervention. This distinction may be also captured in the difference between ``making choices'' and ``executing processes'', with humans being endowed with the power to do the former, and machines only ever being able to do the latter (so far).\footnote{See \cite{wood2026extensions}.} As succinctly put by Heather Roff, ``[m]achine autonomy is therefore fundamentally different than philosophical or moral autonomy. Machine autonomy is merely the ability to act in the world without someone or something immediately directing that action''.\footnote{\cite{roff2013killing}, p. 353.}

% Autonomy comes on a sliding scale, and we should not be viewing automatic, automated, and autonomous as distinctly separate categories, but rather as shorthands for specific areas of the sliding scale of autonomy; no machine system can ``make decisions'', instead they ``execute processes'' (\cite{wood2026extensions}); to that extent, every machine system is ``automatic'' in the sense picked out by authors wishing to draw hard distinctions; this is unhelpful though; it is more useful to simply treat autonomy as a single sliding scale
% No clear technical distinction between automatic, automated, and autonomous (ICRC 2019; see Taddeo and Blanchard citation); some set a very high bar for autonomous, making it akin to human agency; others set a very low bar for autonomous; those setting a high bar would view the low bars as not picking out ``autonomy'', but rather the quality of something being ``automatic'' or ``automated''; many of those setting the high bar and dismissing the low bar simply dress up a low bar in fancy agentic language though (see Taddeo and Blanchard); a simple sliding scale of autonomy with greater permissiveness in planning and execution and less human input at the top, more at the bottom, is most sensible
In the literature on autonomous systems, differing views of autonomy have been put forward, with some focusing on scales of narrow machine autonomy rooted in levels of independence in task or process execution,\footnote{E.g., \cite{yang2017medical,ezenkwu2019machine,haidegger2019autonomy}.} others focusing on more strict requirements such as being ``capable of understanding higher-level intent and direction'' as hurdles to be cleared before being considered ``autonomous''.\footnote{\cite{ukmod2017unmanned}, p. 13; \cite{williams2015defining}, p. 33; \cite{boothby2016weapons}, p. 248. \cite{taddeo2022comparative} attempt to distinguish between ``automatic'' and ``autonomous'' systems in a similar way, but fail to adequately draw this distinction.} A more general expression of autonomy which allows for the determination of differing levels and ranges of capability is given by \cite{antsaklis2020autonomy}, who clarifies that ``autonomy and autonomous imply that the system has the capability to accomplish certain goals'', with autonomy ``always [being] with respect to certain goals'', and an ``autonomous system always [having] a set of goals to be achieved and a control mechanism to achieve them. This implies that \textit{every autonomous system is a control system}''.\footnote{\cite{antsaklis2020autonomy}, pp. 16--17.}

Now, scholars could (and likely will) continuously debate the exact contours of and distinctions between ``automatic'', ``automated'', and ``autonomous'' systems. But given that ``there is no clear technical distinction between automated and autonomous systems, nor is there universal agreement on the meaning of these terms'',\footnote{\cite{icrc2019autonomy}, p. 7.} it is far more useful to treat autonomy as a sliding scale, with merely (mechanistically) automatic entities at one end of the spectrum and full moral agents endowed with ``first mover'' capabilities at the other.\footnote{That is, assuming one allows for first movers in one's ontology.} It should be settled though that machines do not ever ``make decisions'' in a philosophically, ethically, or legally relevant manner, and only ever ``execute processes''. To that extent, all machines are ``automatic'' in the sense picked out by those authors wishing to draw hard distinctions between terms. 

% \textit{In situ} learning is almost certain to rule out the permissible use of an AWS; picking out this class of system may be useful for developing sensible targeted bans already easily drawn from IHL though
A further point worth noting is that while it would be prejudicial to define all AWS as inherently having \textit{in situ} or online live learning (for reasons argued above), this capability can certainly be identified in refined definitions of particular types or classes of AWS. Moreover, such a refined definition is useful for picking out precisely these sorts of systems which likely would be illegal to use, and thus worth pre-emptively banning. 

Regardless of one's preferred views on what constitutes ``autonomy'' in systems, a refined modular conception of particular (classes of) AWS should clearly and explicitly identify what level of autonomy a system has, with the exact contours of autonomy levels or distinctions laid out following the presented definition.\footnote{\cite{williams2015defining}, pp. 39--52 provides useful extensive analysis of levels of autonomy in the military domain, and \cite{antsaklis2020autonomy} gives a mathematical approach of general applicability.} An example refinement may look as follows:

\begin{quote}
	\textbf{Autonomous Weapon Systems (refined definition with autonomy):} Machine-based weapon systems [with autonomy of level $n$] that, if properly functioning, once activated, can identify, select, and engage targets without contemporaneous human input.
\end{quote}

\subsection{F2T2EA}\label{subsec_conceptual-framework_f2t2ea}

% Multiple modularity around the find, fix, track, target, engage, assess kill chain loop
Related to the above, it is also critical to recognize that individual systems may have differing levels of autonomy in different tasks or areas of functioning. We should therefore be ``thinking about `autonomous functions' of systems, rather than characterizing an entire vehicle or system as `autonomous'\thinspace'',\footnote{\cite{scharre2015introduction}, p. 7.} especially as autonomy may be used in isolated ways for an incredibly wide array of tasks, from streamlining mundane back-office aspects of military institutions to final kinetic engagement of enemy units.\footnote{See \cite{meerveld2023irresponsibility,meerveld2025operationalising,knack2025defence,voloshchuk2025artificial,wood2026stop}. Cf. \cite{dubber2026domain} for limits to this.} One way to conceptualize the distinct tasks or functions augmented by autonomous capabilities is to look at the observe-orient-decide-act decision loop (OODA Loop) developed by U.S. Air Force Colonel John Boyd,\footnote{\cite{enck2012ooda}.} or to examine the more fine-grained breakdown of so-called ``kill chains'' through the F2T2EA model (find, fix, track, target, engage, assess).\footnote{\cite{tirpak2000find}.}

% POSSIBLY ADD GRAPHIC/CHART HERE OF THE MODERN KILL CHAIN

Considering the latter, each phase of the F2T2EA model may involve more discrete potentially autonomous capabilities as well. For example, the ``find'' phase could involve systems which autonomously gather intelligence data, autonomously assess that data, autonomously flag objects/persons considered to be hostile, or some combination of these. Adding autonomy at different phases or for different functions can also implicate different ethical, legal, technical, or operational concerns. For example, autonomously tracking a target which a human has previously identified based on their own situational awareness is much less fraught than having a system that autonomously finds a target and then hands off tracking to a human agent. Full autonomy in engagement and assessment also raise numerous concerns not present with autonomous functions at other phases of the F2T2EA model. In short, it is not just the level of autonomy that matters, but also for what that autonomy is used. For refined definitions, this calls for an explicit mention of where autonomy factors into some system under consideration, both in terms of the functions where autonomy is present and the uses to which that system is put. 

\begin{quote}
	\textbf{Autonomous Weapon Systems (refined definition with autonomy, functions):} Machine-based weapon systems [with autonomy of level $n$] in [functions $x,y,z$] that, if properly functioning, once activated, can [execute functions $x,y,z$] without contemporaneous human input.
\end{quote}

\subsection{Lethality}\label{subsec_conceptual-framework_lethality}

% Highlight that while there are serious problems with this category, given the centrality of it in debates and its prominence in regulatory discussions (see GGE rolling text), it is worth including, and likely as one of the first categories; clarify though that this is a poor category to include, as virtually any weapon can be used in lethal ways, and some ``non-lethal'' systems can through user error cause accidental fatalities
% Raise the more standard view of examining ``less-lethal'' systems and perhaps ``lethal by design'' systems
In many discussions of AWS, lethality is a central element driving debate.\footnote{E.g., \cite{gge2026reportrev3,ccwgge2026ngoreport}.} %, a prominent example being the work of the long-standing Group of Governmental Experts (GGE) on Lethal Autonomous Weapon Systems (LAWS) operating under the auspices of the Convention on Certain Conventional Weapons (CCW).\footnote{For the rolling text of the GGE's efforts at establishing regulation, see \cite{gge2025rolling12052025}.}
Now, there are fundamental issues with this category, given that lethality is generally a characteristic of use, and not of systems \textit{per se}. Thus, a household hammer can be lethal or non-lethal in virtue of use, and there are few design decisions that could rule out its lethality. By the same token, high-explosive ordnance may be non-lethal, if it is directed purely at materiel and objects and not used against personnel or manned platforms. It is thus more correct to speak of systems \textit{designed to be lethal} and those \textit{designed to be less lethal} (as non-lethality generally cannot be guaranteed).\footnote{\cite{kobayashi2009rubber}; \cite{boothby2016weapons}, pp. 233--237. See \cite{wood2020problem} for arguments against designing AWS to be lethal.} The only categories where lethality might be conclusively ruled out is for AWS operating purely in the electromagnetic environment (see Section \ref{subsec_conceptual-framework_ew}), in the cyber domain (see Section \ref{subsec_conceptual-framework_cyber}), or in information/cognitive warfare systems (see Section \ref{subsec_conceptual-framework_io-cognitive}). 

% Close by highlighting that some weapons can be reasonably treated as ``non-lethal'', as their functioning makes it highly unlikely that one could kill with them (barring intentional use of the system far outside its normal parameters); e.g., some microwave weapons, flashbangs, bio-chem weapons which can only foreseeably incapacitate (though note the legal issues with these from BWC/CWC); cite discussion from \cite{boothby2016weapons}, pp. 233--237
The problems with lethality as a definitional marker notwithstanding, this is a characteristic regularly raised in debates on AWS, and its very close relationship with anti-personnel AWS also makes it a worthwhile feature to add. Any addition of this feature should highlight though that more/less lethality is as much derived from system use as it is from system design. It should also indicate that \textit{lethal} AWS, that is, LAWS, are generally only to be LAWS because they \textit{are designed to be lethal}, and not simply because they \textit{could be lethal in use}. 

\begin{quote}
	\textbf{Autonomous Weapon Systems (refined definition with autonomy, lethality):} Machine-based weapon systems [with autonomy of level $n$] and [designed to be lethal/less-lethal] that, if properly functioning, once activated, can identify, select, and engage targets without contemporaneous human input, [resulting in death/not expected to result in death].
\end{quote}

\subsection{Anti-Personnel/Anti-Materiel}\label{subsec_conceptual-framework_antipersonnel-antimateriel}

% In line with above distinction, draw out that a core distinction will be whether the core functionality of a system is to counter persons or objects; clarify that anti-materiel systems may also be lethal, potentially by design, in virtue of being designed to destroy manned systems
As ``lethality'' is related just as much to target classes (personnel vs. materiel) as it is to system design, this is a further core element which may feature in refined definitions. Whether a system is designed to target human beings also raises numerous ethical and legal worries, and poses unique engineering challenges as well. Anti-materiel systems, on the other hand, may often be designed so as to rely on very distinct markers of some platform type, and their ability to be deployed against platforms exclusively, potentially even platforms devoid of humans (incoming missiles, ordnance, or unmanned vehicles), makes their deployment less fraught and the need for regulation less acute. There are thus deep governance reasons to clearly distinguish the anti-personnel/anti-materiel character of certain AWS, should they be designed for one or another target class exclusively. Notably, purely anti-materiel systems may also be lethal, potentially by design, in virtue of being designed to destroy manned platforms, but the fact that they are by necessity targeting platforms greatly alters (and simplifies) the moral and legal considerations being made.\footnote{Technological and doctrinal considerations are likewise likely to be improved, given that machine methods for identifying and verifying platforms are generally more reliable than those for assessing humans and their combatancy.}

\begin{quote}
	\textbf{Autonomous Weapon Systems (refined definition with autonomy, functions, target class):} Machine-based [anti-personnel/anti-materiel] weapon systems [with autonomy of level $n$] in [functions $x,y,z$] that, if properly functioning, once activated, can [execute functions $x,y,z$] against [individuals/objects] without contemporaneous human input.
\end{quote}

\subsection{Locomotion and Offensive/Defensive Systems}\label{subsec_conceptual-framework_locomotion}

% Under this heading, we can also subsume the offense/defense distinction; clarify that there are challenges to this as a hard distinction, but that it can be a useful heuristic for classifying some sorts of weapons and their attendant challenges (i.e., a CIWS faces very distinct challenges from a counter-batter AWS, though both may be ``preventing incoming fires'')
The offensive/defensive character of a system and whether it can move can also be relevant to regulatory and doctrinal discussions. While it is generally difficult to draw clean distinctions between ``offensive'' and ``defensive'' weapon systems, as use will impact on this greatly, a system's ability to move (or not) can significantly alter its effective ability to be deployed in offensive or defensive roles. Given that defensive uses of force are generally easier to justify and more likely to generate greater latitude in decision-making, %(because fewer alternatives are available to the defending party), 
this factor may be relevant to build into certain refined definitions (based on use case). Thus, while there may be difficulties in establishing hard distinctions between offensive and defensive weapons, the terms can offer useful heuristics for debate, and locomotion represents a core capability of offensive systems. Where AWS are fixed to points or other platforms, they will often be serving in defensive roles for those points/platforms, and both their static nature and greater likelihood to fulfill a defensive role may impact on regulation or doctrine. 

\subsection{Swarmed}\label{subsec_conceptual-framework_swarm}

While not yet a fully mature class of technology, there is increasing interest in swarming capabilities for AWS, especially aerial systems which may be designed to achieve mission success through saturation rather than precision and hard-to-intercept fires. Developers claim continuing progress in this domain,\footnote{E.g., the KITU2 program of Quantum Systems, which published milestone achievements in 2024. See \cite{quantumsystems2024swarmprogress,quantumsystems2024swarmmilestone}.} and the trend in unmanned and autonomous aerial systems is toward greater platform integration and swarming capabilities. Swarming raises distinct ethical, legal, technical, and operational concerns though, and AWS which can execute swarming operations or designed to be swarmed systems should have that element picked out in refined definitions which drive debate around swarms/swarming. 

\subsection{Kinetic/Explosive}\label{subsec_conceptual-framework_kinetic}

Systems which engage targets using kinetic or explosive force raise distinct challenges relating to whether they would ``cause superfluous injury or unnecessary suffering''\footnote{AP I, Art. 35.2.} and regarding the risks they pose to civilians. In particular, given that combatants are required to distinguish between civilians and combatants, directing attacks only at the latter;\footnote{AP I, Art. 48.} that combatants must use methods and means of warfare which can be directed at military objectives;\footnote{AP I, Art. 51} and that combatants must take all feasible precautions to minimize harm to civilians and warn them of risks attending attacks,\footnote{AP I, Art. 57.} kinetic/explosive methods and means of war create special challenges. This is because attacks causing kinetic/explosive effects may create immediate and potentially lethal risks to the civilian population. Picking out the (non-)kinetic/explosive character of AWS in refined definitions may thus be relevant to specific debates and potential regulatory efforts, especially those relating to the use of such weapons in urban environments.\footnote{For example, the Political Declaration on Strengthening the Protection of Civilians from the Humanitarian Consequences Arising from the Use of Explosive Weapons in Populated Areas (\url{https://www.gov.ie/en/department-of-foreign-affairs/publications/protecting-civilians-in-urban-warfare/}).}

\subsection{Electronic Warfare}\label{subsec_conceptual-framework_ew}

Electronic warfare (EW) has become increasingly critical in modern war,\footnote{For core early and contemporary work on EW, see, respectively, \cite{vakin2001fundamentals,zhang2023radar,yu2025electronic}.} and emerging AWS, especially those designed to counter unmanned and autonomous systems, may increasingly feature EW capabilities or even be given EW roles as their primary task.\footnote{\cite{burton2019autonomous,sharma2020artificial,casterline2022applications}.} Where kinetic attacks may directly cause risks of physical harm for combatants and civilians, EW interferes with the electromagnetic spectrum in which communications and data are transmitted, and through which core elements of navigation are relayed. This thus creates equally real risks for individuals, but mediated through failures of navigation, identification of (air)craft, or related concerns. Much EW is also difficult or impossible to tightly direct toward purely military objects, raising issues of distinction.\footnote{See AP I, Art. 51.} Given these distinct issues germane to EW, but less relevant or even nonexistent for other platforms, this may be a capability in need of explicit mention in refined definitions focusing on regulation around EW at large, and AWS capable of engaging in EW.

\subsection{Cyber}\label{subsec_conceptual-framework_cyber}

Like EW, cyber-warfare does not immediately imply risks of harm to either combatants or civilians, but it can indeed lead to harm through the effects of cyber-attacks. Cyber-attacks can target both software and the hardware upon which software systems are built\footnote{\cite{mantellassi2024cloudreport,bray2026hardware}.} and can be used to target civilian infrastructures\footnote{\cite{tsakanyan2025modern}.} or feature as a core element of kinetic warfighting.\footnote{\cite{kajander2025cybersecurity,johansmeyer2026small}.} The dual-use nature of many critical infrastructures and the difficulty in conducting cyber-attacks that target purely military systems raises significant legal worries as well.\footnote{\cite{boothby2014where,midson2014geography,schmitt2012attack,schmitt2017tallinn}.} Furthermore, some forms of cyber-attack can be difficult to control once initiated, raising legal concerns over distinction and one's ability to control the effects of attacks (as required by AP I, Art. 51.4.c).\footnote{\cite{dubber2025military}.} For these reasons, AWS which are designed to execute attacks or carry out tasks in the cyber domain raise distinct worries and are worth having their cyber capabilities picked out in refined definitions dealing with the intersection of AWS and cyber-warfare. 

\subsection{Information/Cognitive Warfare}\label{subsec_conceptual-framework_io-cognitive}

In addition to systems which are clearly weapons, some tactics of warfighting have nothing to do with exerting force on a battlefield, either directly or indirectly, and instead target the will and resolve of a population. Information operations can be carried out to sow uncertainty, entrench and exacerbate societal divisions, muddy effective response to aggression, or win over part of a population to an aggressor's cause (to name just a few).\footnote{For general reading on information operations, see, e.g., \cite{gioe2020soviet,tatham2024information,olejnik2025propaganda,havlik2026war}. For ethical implications, see \cite{floridi2014ethics,taddeo2014information}.} Cognitive warfare, on the other hand, can be leveraged to disrupt and undermine the very ways that a population thinks, making them untrusting of information, causing them to withdraw from complex issues, or exerting ``reflexive control'' to ``prompt behavior that will give the principal party a competitive advantage''.\footnote{\cite{degoeij2023reflexive}, p. 99. See also \cite{hoffman2025assessing,kallioniemi2025beyond,zilincik2026cognitive}.} One may object to these being brought under the umbrella of autonomous weapon systems, but the increasing capability of large language models and agentic AI systems is allowing for information and cognitive warfare to be massively scaled up with very little cost,\footnote{\cite{bukowski2025agents,krewson2025brain,russo2026cognitive}.} and using fully autonomous systems which can directly engage a target population without direct human input or oversight.\footnote{\cite{hartmann2020next,goldstein2023generative,fredheim2024assessing}.} 

This is not to say that a definition of AWS need be so broad as to include these psychological aspects of modern warfighting, but to highlight that a modular definitional schema can accommodate such expansion and specificity as needed, or can be used to pick out more traditional kinetic systems with ease. In any event, the approach developed here creates clear, traceable, and useful routes for introducing or eliminating components from a particular definition, greatly improving over the status quo alternatives. 

\subsection{General Recommendations on Definition Construction}\label{subsec_conceptual-framework_general-recommendations}

As noted by \cite{dubber2026abhorrence}, ``military autonomy spans domains (air, land, sea, space, cyber), functions (targeting, defence, logistics, intelligence, command support), levels (tactical, operational, strategic), and phases of conflict (cooperation, competition, crisis, conflict)'' and ``[e]ach combination of these dimensions presents a distinct profile of control, error, reversibility, and moral risk---and treating them as a single class obscures more than it reveals.''.\footnote{\cite{dubber2026abhorrence}, p. 12.} A unary conception of AWS is thus unfit for any but the most broad-based and fundamental of tasks, such as establishing basic understandings and grounding principles. For anything more specific -- and there is much need for more specific rules and best practices in a host of use cases -- refined definitions will be necessary. In developing such refinements, scholars, practitioners, and policy-makers should focus first and foremost on the use case they have in mind which requires a definition. As no single view is likely to be fully satisfactory, the central concern should be whether definitions are \textit{fit for purpose}. 

In crafting purpose-driven definitions, it should also be clear that various terms may be slotted into the base definition in different manners, as illustrated in Sections \ref{subsec_conceptual-framework_autonomy}--\ref{subsec_conceptual-framework_antipersonnel-antimateriel}. There will not be one ``correct'' way to incorporate relevant terms in refined definitions, as the exact reason for having a definition and the stakeholder group who will use it can greatly impact on how best to develop a refined understanding. No matter how one goes about this though, it is centrally important that terms be viewed in a pragmatic fashion, as modules which can be leveraged in a ``plug-and-play'' fashion to serve discrete regulatory, doctrinal, or scholarly needs. 

As argued at the outset of this article, man-made artifacts do not readily conform to a ``scientific'' definition, and in domains where the exact definition of a term can greatly color subsequent discourse, there is value in being agnostic and pragmatic about how precisely to define core terms. That is not to say that definitions do not matter or may be constructed simply as one sees fit. But recognizing the diversity of views on AWS (or autonomous systems, artificial intelligence, etc.), we should retain both enough intellectual humility to recognize our preferred definition may be reasonably objected to and enough humanistic feeling to put progress ahead of ``being correct'' on definitional matters. 

This means that constructing definitions should be a group exercise carried out by relevant stakeholders in a domain. Such stakeholders should be open and honest about what matters to them in a precise definition, what concerns they have which they believe may be better discussed with a shared definition in mind, and what political and other aims they wish to pursue using that definition (as honesty on that front can prevent unnecessary confusion and wasted time). 

\section{Conclusion}\label{sec_conclusion}

It has been over ten years since the Group of Governmental Experts on Lethal Autonomous Weapon Systems began its discussions, and still there is no broadly agreed consensus in the international community on how best to regulate AWS, nor even on what AWS are.\footnote{Cf. though the recent progress of the GGE, which, though not yet universally agreed upon, may set a baseline for LAWS relevant to AWS as well. See \cite{gge2026reportrev3}.} There are many reasons for this glacial and uncertain progress, but recurrent debates and relitigations of definitions has certainly not helped matters. Moreover, certain actors have added mud to the definitional waters simply for the sake of political ends,\footnote{E.g., \cite{roff2024magnifying}, discussing the early debates on ``meaningful human control'', a term which was ``a `political concept' designed to get States to a negotiating table and engaging in discussion''.} with resulting intellectual debates faltering on core terms while technologies continue evolving with little explicit oversight. 

As there is scant hope in trying to simply impose a definition on all parties and being met with agreement, and due to core limitations in any definition treated as \textit{the conception} of AWS, we have argued that we should instead aim for a base definition to be refined with core terms as needed. Such a definitional schematic framework allows one to remain open to changes in technology, to fit definitions to discrete needs and systems, and to develop distinct definitions suited to different stakeholder groups. The guiding light in this approach is the blunt pragmatism embodied by the International Committee of the Red Cross, which wisely recalls that ``unable to eliminate the scourge of war, one endeavors to master it and mitigate its effects''.\footnote{\cite{icrc1987ap1commentary}, p. 480, at paragraph 1601.} The point is not to have the ``right'' view. It is to have a view which is useful in helping us to master the scourge of war, to mitigate its effects. 

\pagebreak 
%{\small \bibliography{D:/Files/Dropbox/School/master}}
{\small \bibliography{master}}

\end{document}